\documentstyle[12pt,a4]{article}

\font\mybb=msbm10 at 12pt
\def\bb#1{\hbox{\mybb#1}}

\def\R {\bb{R}}
\def\C {\bb{C}}
\def\J {\bb{J}}
\def\I {\bb{I}}

\def\ele{\mathop{\rm ele}\nolimits}
\def\mag{\mathop{\rm mag}\nolimits}
\def\tr{\mathop{\rm tr}\nolimits}
\def\Im{\mathop{\rm Im}\nolimits}
\def\diag{\mathop{\rm diag}\nolimits}
\def\rank{\mathop{\rm rank}\nolimits}
\def\Tr{\mathop{\rm Tr}\nolimits}
\def\mod{\mathop{\rm mod}\nolimits}

\makeatletter
\@addtoreset{equation}{section}
\@addtoreset{footnote}{page}
\makeatother

\newcommand{\vs}{\vspace*}
\newcommand{\hs}{\hspace*}
\newcommand{\wt}{\widetilde}
\newcommand{\ol}{\overline}
\newcommand{\ra}{\rightarrow}
\newcommand{\lra}{\leftrightarrow}
\newcommand{\nn}{\nonumber}
\newcommand{\mms}[1]{\makebox[4ex]{$#1$}}
\newcommand{\sq}{\sqrt{2}\,}
\newcommand{\VEV}[1]{\left\langle #1\right\rangle}

\begin{document}
\begin{titlepage}

\begin{flushright}
KUNS-1451\\
HE(TH)97/10\\
hep-th/9705069
\\
May, 1997
\end{flushright}

\vs{6ex}
\begin{center}
{\large \bf
Deformations of ${\cal N}=2$ Dualities to ${\cal N}=1$ Dualities
in $SU$, $SO$ and $USp$ Gauge Theories
}
\vspace{12ex}

Takayuki Hirayama
\footnote{e-mail: hirayama@gauge.scphys.kyoto-u.ac.jp},
Nobuhiro Maekawa
\footnote{e-mail: maekawa@gauge.scphys.kyoto-u.ac.jp}
and
 Shigeki Sugimoto
\footnote{e-mail: sugimoto@gauge.scphys.kyoto-u.ac.jp}
\vspace{0.5cm}

{\it Department of Physics, Kyoto University,\\
     Kyoto 606-01, Japan}
\end{center}
\vspace{1cm}

\vs{6ex}
\begin{abstract}
We study deformations of dualities in finite ${\cal N}=2$ supersymmetric QCD.
Adding mass terms for some quarks and the adjoint matter
to the finite ${\cal N}=2$ theory, which is known to have dual descriptions,
the correspondence of gauge invariant operators between
the original and dual theory is deformed. 
As a result, we naturally obtain N.Seiberg's ${\cal N}=1$ duality.
Furthermore, we discuss the origin of the meson and superpotential in the dual
theory. This approach can be applied to $SU(N)$, $SO(N)$, and $USp(2N)$ 
gauge theories, and we analyze all these cases.
\end{abstract}


\end{titlepage}

\section{Introduction}
In the last few years remarkable progress has been made in
supersymmetric gauge
theories in four dimensions. Using the methods developed by N.Seiberg,
effective superpotential can often be determined exactly and lots of
non-perturbative effects are found (see \cite{lec} for a review).
One of the most interesting discoveries is duality in ${\cal N}=1$
supersymmetric gauge theories. For example,
in the pioneered work of N.Seiberg \cite{S}, it was found that
 ${\cal N}=1$ supersymmetric $SU(N_c)$ QCD with $N_f$ flavors of quarks 
has a dual description, which is ${\cal N}=1$ supersymmetric $SU(N_f-N_c)$
QCD with $N_f$ flavors of dual quarks and a gauge singlet meson field
interacting
with the dual quarks by the superpotential.

Up to now there is no rigorous proof of this duality, 
and so it is important to investigate non-trivial evidences for it.
There are several arguments that the ${\cal N}=1$ duality can be derived
by analyzing ${\cal N}=2$ supersymmetric QCD \cite{APS,APSh,LS}.
Since N.Seiberg and E.Witten's two beautiful papers \cite{SW} appeared,
it has become clear that the low energy effective theory of ${\cal N}=2$
supersymmetric QCD can be analyzed exactly making use of hyperelliptic
curves \cite{APSh1,HO}. In \cite{APS}, it was shown that the low energy
effective theory at the baryonic root of ${\cal N}=2$ supersymmetric $SU(N_c)$
QCD with $N_f$ flavors is $SU(N_f-N_c)\times U(1)^{2N_c-N_f}$ gauge theory,
and adding ${\cal N}=2$ breaking mass term for the adjoint chiral field,
the theory flows to ${\cal N}=1$ supersymmetric $SU(N_f-N_c)$ QCD which
is consistent to the N.Seiberg's duality in ${\cal N}=1$ supersymmetric QCD.
This argument can also be applied to $SO(N)$ and $USp(2N)$
\footnote{$USp(2N_c)$ is the unitary symplectic group of rank $N_c$.}
 gauge theory
\cite{APSh}. However, these works do not provide a complete derivation 
of the N.Seiberg's duality since the origin of 
the meson field and the superpotential, needed in the
dual theory, is not specified. An approach for this problem is given
in \cite{LS}, in which finite ${\cal N}=2$ supersymmetric QCD is
considered in detail. ${\cal N}=2$ supersymmetric $SU(N_c)$ QCD with $2N_c$
flavors is known to be finite and believed to have a dual description.
Adding mass term
for the adjoint chiral field to this theory and using Fierz
transformation, the authors of \cite{LS} derived the dual $SU(N_c)$
theory with the meson field and the superpotential.

The purpose of this paper is to obtain deeper understanding of
the relation between  ${\cal N}=1$ duality and 
${\cal N}=2$ duality, and to derive ${\cal N}=1$ duality
from ${\cal N}=2$ duality.
In particular, we generalize the argument in \cite{LS} for the case
with $N_f$($\leq 2N_c$) massless flavors and propose the origin of the
meson field and the superpotential.

We start with the finite ${\cal N}=2$ theory
with mass terms for hypermultiplets.
The dual of this theory is determined so as to have the
same hyperelliptic curve \cite{APSh1}.
Then we add ${\cal N}=2$
breaking mass term for the adjoint chiral field to obtain ${\cal N}=1$
theory and see the change of the vacuum moduli space. This change of
the vacuum
moduli space is non-trivial because we
must correctly consider non-perturbative effects. To have the same
vacuum moduli space, we determine
corresponding deformation for the dual theory and the correspondence 
of gauge invariant operators between the original and dual theory.
Now we can apply the transformation used in \cite{LS} and 
 obtain N.Seiberg's ${\cal N}=1$ dual theory including a meson field
interacting with the dual quarks by the superpotential.
This method can be applied to $SU(N_c)$, $SO(N_c)$ and $USp(2N_c)$ gauge
theories. In the case of $SO(N_c)$ and $USp(2N_c)$ gauge theories,
unlike the $SU(N_c)$ case,
we find that the vacuum moduli space has several distinct branches, and the
dual transformation maps each branch to the corresponding dual branch.

\section{S-duality in ${\cal N}=2$ $SU(N_c)$ Gauge Theory}

\subsection{a brief review of S-duality in ${\cal N}=2$ theory}
\label{S-duality}
In this subsection we will make a brief review of S-dualities in ${\cal N}=2$
supersymmetric QCD. Here we consider ${\cal N}=2$ 
supersymmetric $SU(N_c)$
QCD with $N_f$ hypermultiplets in the ${\bf N_c}$
representation of the gauge group.
The theory can be described in terms of ${\cal N}=1$
superfields: $W_\alpha$ (a field strength chiral multiplet), $\Phi$
(a chiral multiplet in the adjoint representation 
of the gauge group), $Q^i$ and
$\wt{Q}_i$ (chiral multiplets in the ${\bf N_c}$ and ${\bf \ol{N_c}}$
representation of the gauge group respectively),
where $i=1,\cdots,N_f$ are flavor
indices. The superpotential is
\begin{eqnarray}
  W_{\ele}= \sq gQ^i\Phi \wt{Q}_i +\sq m_i^j Q^i\wt{Q}_j,
\label{Wele}
\end{eqnarray}
where $m=(m_i^j)=\diag(m_1,\cdots,m_{N_f})$ is a quark mass matrix.

When $N_f<2N_c$, the theory is asymptotically free and when
$N_f=2N_c$, the theory is scale-invariant (for $m=0$).
Highly nontrivial evidences of S-duality have been found for
$N_f=2N_c$ and so we restrict our attentions to this case.

The vacuum moduli space of the theory was analyzed in detail
in \cite{APSh1,APS}. They showed that the vacuum moduli space has various
branches intersecting with each other, and is locally a product of 
a Coulomb branch and a Higgs branch
\footnote{Due to the non-renormalization theorem, the local product
structure is retained quantum mechanically.}.
 The Coulomb branch can be analyzed exactly making use of the hyperelliptic
curves derived in \cite{APSh1,HO}:
\begin{equation}
  y^2=\prod_{a=1}^{N_c}(x-\phi_a)^2+4h(h+1)\prod_{i=1}^{N_f}(x-m_i-2m_S h),
  ~~~N_f=2N_c
\end{equation}
where $m_S\equiv (1/N_f)\sum m_i$ is the flavor-singlet mass,
$h(\tau)\equiv\theta_1^4(\tau)/(\theta_2^4(\tau)-\theta_1^4(\tau))$
is a specific modular function of the bare gauge coupling constant
$\tau=\theta/\pi+i8\pi/g^2$.
This curve is invariant under S-duality transformation 
$\tau\ra -1/\tau$, $m_i\ra\wt{m_i}\equiv m_i-2m_S$. So it is strongly
suggested
that there is another description of the same physics. We want to call
the original and dual description by the electric and magnetic theory
respectively. The magnetic theory is also ${\cal N}=2$ supersymmetric
$SU(N_c)$ QCD with $N_f$ hypermultiplets in the ${\bf N_c}$
representation of the gauge group, but the bare masses and the
couplings are different.
The superpotential is
\begin{eqnarray}
  W_{\mag}=\sq\wt{g}q_i\varphi \wt{q}^i
  +\sq\wt{m}_i^jq_j\wt{q}^i, \label{Wmag}
\end{eqnarray}
where $\varphi$, $q_j$ and $\wt{q}^i$ are chiral multiplets
\footnote{Note that we use the notation in \cite{S} for $q_i$ and $\wt{q}^i$,
which is different from that in \cite{APS}.}
in the adjoint, ${\bf N_c}$ and ${\bf\ol{N_c}}$ representation of the 
gauge group respectively, $\wt{m}_i^j\equiv m_i^j-2m_S\delta_i^j$
and $\wt{g}\equiv 1/g$.

The Higgs branches do not receive quantum corrections and 
can be analyzed classically, due to the non-renormalization theorem
\cite{APS}.
They are parameterized by holomorphic gauge invariant operators with several
constraints \cite{LT}. 
As a result of detailed analysis in \cite{APS},
it was shown
that there is a correspondence of gauge invariant operators 
between the electric and magnetic theory,
which is compatible with all the constraints.
Namely, the Higgs branches are also the same in the electric and
magnetic theory.

These facts strongly suggest that the electric and magnetic theory describe
the same physics. And more surprisingly, adding adjoint mass terms,
this ${\cal N}=2$ duality flows down to
the duality of ${\cal N}=1$ supersymmetric QCD found by N.Seiberg \cite{S}.
We examine this point in the next section.

All these analysis are applicable to $SO(N_c)$ and $USp(2N_c)$
gauge theories.
We investigate these cases in section \ref{SO} and section \ref{USp}.

\subsection{the correspondence of the Higgs branches}
\label{N2duality}
Let us explain in some detail the electric-magnetic
correspondence of gauge invariant operators
in the Higgs branches of the ${\cal N}=2$ supersymmetric $SU(N_c)$ QCD
with $2N_c$ flavors. Using the same argument as in \cite{APS}, we can
determine the structure of the Higgs branches for the case there are
mass terms for the hypermultiplets, and show that there is a
electric-magnetic correspondence of gauge invariant operators.

In the electric theory, F-term equations are as follows.
\begin{eqnarray}
  Q^i_a\wt{Q}_i^b&=&\rho\delta_a^b,~~~(\rho\in \C),  \label{F1}  \\
  g \Phi^a_b\wt{Q}_j^b+m^i_j \wt{Q}_i^a&=&0,  \label{F2}  \\
  g Q^i_a\Phi^a_b +m^i_j Q^j_b&=&0.  \label{F3}
\end{eqnarray}
We denote the vacuum expectation values of
$\Phi$, $Q$ and $\wt{Q}$ by the same symbols.
(\ref{F2}) and (\ref{F3}) imply that
we do not need the gauge invariant operators
which are the mixture of $\Phi$ and ($Q^i,\wt{Q}_i$) when we describe
the moduli by the gauge invariant operators.
The moduli space, which consists of the Coulomb branch and the Higgs
branches, is parameterized by $U_k\equiv \tr(\Phi^k)$,
$(k=2,\cdots,N_c)$, the meson and the baryons. Roughly speaking,
the Coulomb branch is parameterized by $U_k$
and the Higgs branches are parameterized by the meson and the baryons.

The meson and the baryons are defined as
\begin{eqnarray}
  M^i_j&\equiv&Q^i_a\wt{Q}_j^a, \\
  B^{i_1\cdots i_{N_c}} &\equiv&
  Q^{i_1}_{a_1}\cdots Q^{i_{N_c}}_{a_{N_c}}\epsilon^{a_1\cdots a_{N_c}}, \\
  \wt{B}_{i_1\cdots i_{N_c}} &\equiv&
  \wt{Q}_{i_1}^{a_1}\cdots \wt{Q}_{i_{N_c}}^{a_{N_c}}
  \epsilon_{a_1\cdots a_{N_c}}. 
\end{eqnarray}
By definition, the meson and the baryons are subjected to the constraints:
\begin{eqnarray}
  (*B)\wt{B}&=&*(M^{N_c}),  \label{BBM} \\
  (*B)\cdot M&=&M\cdot *\wt{B}=0,  \label{BM0} \\
  (*B)\cdot B&=&\wt{B}\cdot *\wt{B}=0.  \label{BB0}
\end{eqnarray}
Here we use the notations in \cite{APS} that the ``$\cdot$'' represents the
contraction of a lower with an upper flavor index, and the ``$*$'' stands for
contracting all flavor indices with the totally antisymmetric tensors
$\epsilon_{i_1\cdots i_{2N_c}}$ or $\epsilon^{i_1\cdots i_{2N_c}}$.
For example (\ref{BBM}) is
\begin{equation}
\epsilon_{i_1\cdots i_{N_c}k_1\cdots k_{N_c}}
 B^{k_1\cdots k_{N_c}} \wt{B}_{j_1\cdots j_{N_c}}
=\epsilon_{i_1\cdots i_{N_c}k_1\cdots k_{N_c}}
M^{k_1}_{j_1}\cdots M^{k_{N_c}}_{j_{N_c}}.
\end{equation}
There are further constraints for the meson and the baryons
from the F-term equations (\ref{F1})$\sim$(\ref{F3}) :
\begin{eqnarray}
  M\cdot M'&=&0,  \label{MM0} \\
  M'\cdot B&=&\wt{B}\cdot M'=0,  \label{MB0} \\
  **(m\cdot B)&=&*(m\cdot \wt{B})=0, \label{mB0} \\
  m\cdot M&=&M\cdot m, \label{mMMm} \\
  m^i_jM^j_i&=&0,  \label{TrmM}
\end{eqnarray}
where
$(M')^i_j\equiv M^i_j-\frac{1}{N_c}(\Tr M)\delta^i_j$
\footnote{We use ``$\Tr$'' for summing up flavor indices, while
  ``$\tr$'' for color indices.}.
Note that (\ref{mB0}) can be rewritten
  in the more useful way $m\cdot B=m\cdot\wt{B}=0$ for generic choice
  of bare masses.

The constraints including $\Phi$ are as follows
\footnote{
  We have assumed that the bare masses are chosen to be generic.}.
\begin{eqnarray}
  S_kB=S_k\wt{B}&=&0~~~(k=1,\cdots,N_c),
  \label{SkB0} \\
  S_kM^{[i_1}_{j_1}\cdots M^{i_l]}_{j_l}&=&0~~~(k+l> N_c),
  \label{SkMl}
\end{eqnarray}
where we have defined $S_k\equiv \epsilon_{a_1\cdots a_{N_c}}
\epsilon^{b_1\cdots b_{k}a_{k+1}\cdots a_{N_c}}
\Phi^{a_1}_{b_1}\cdots\Phi^{a_k}_{b_k}$ instead of $U_k$.
To show (\ref{SkMl}), we used the equation $m\cdot M=0$ which will
be deduced later from the explicit form of $M$, as well as the usual
formula $\epsilon^{a_1\cdots a_{N}}\epsilon_{b_1\cdots b_N}
=\delta^{[a_1}_{b_1}\cdots\delta^{a_N]}_{b_N}$.
These constraints (\ref{BBM})$\sim$(\ref{SkMl}) form a complete set of
the classical constraints. 
Owing to the non-renormalization theorem in ${\cal N}=2$ supersymmetric
gauge theory, we know that the Higgs branches do not receive quantum
corrections \cite{APS}, and so the classical constraints
(\ref{BBM})$\sim$(\ref{TrmM}),
which are the constraints for the Higgs branches,
 are correct quantum mechanically.

Let us solve these equations and determine the structure of the Higgs
branches.
We set $m=\diag(0,\cdots,0,m_{N_f+1},\cdots,m_{2N_c})$ where $N_f$
is the number of massless flavors and $m_i$'s are chosen to be generic.
Then (\ref{mMMm}) imply that $M$ can be put into the form
\begin{equation}
  \left(
    \begin{array}{cc}
      A & 0  \\
      0 & d
    \end{array}
  \right),
\end{equation}
where $X$ is an $N_f\times N_f$ block and
$d=\diag(d_{N_f+1},\cdots,d_{2N_c})$.
Assuming that $\rank X=r$ and $\rank d=s$
\footnote{(\ref{BBM}) and (\ref{BM0}) imply 
  $\rank(M)=r+s\leq N_c$.},
$M$ can be reduced to the following
form up to $SU(N_f)$ massless flavor symmetry and massive flavor permutations.
\begin{equation}
g^2M=
  \left(
    \begin{array}{cccccc}
      X & Y &         &      &     & \\
      0 & 0 &         &      &     & \\
      &   &\mms{d_{N_f+1}}&      &     & \\
      &   &         &\ddots&     & \\
      &   &         &      &\mms{d_{N_f+s}}& \\
      &   &         &      &     &0
    \end{array}
  \right),
  \label{M1}
\end{equation}
where $X$ is an $r\times r$ block and $Y$ is an $r\times(N_f-r)$
block with $\rank(XY)=r$, and $d_i\neq 0$ for $(i=N_f+1,\cdots,N_f+s)$.
Then (\ref{MM0}) gives $(X-\frac{g^2}{N_c}\Tr M)\cdot (XY)=0$ and 
$(d_i-\frac{g^2}{N_c}\Tr M) d_i=0$, implying 
\begin{eqnarray}
  X^i_j&=&\frac{1}{N_c}(\Tr X+\sum_{k=1}^s d_{N_f+k})\delta^i_j, 
  \label{Adiag}
  \\
  d_i&=&\frac{1}{N_c}(\Tr X+\sum_{k=1}^s d_{N_f+k}).
  \label{d_i}
\end{eqnarray}
The solutions of these equations exist for $r+s=N_c$ or 
$X=s=0$.

First we consider the case $\Tr M\neq 0$, which implies $r+s=N_c$.
The constraint (\ref{TrmM}) gives
\begin{equation}
  \Tr M \sum_{k=1}^{s}m_{N_f+k}=0.
\end{equation}
For the generic choice of bare masses, this equation implies $s=0$.
$Y$ can be taken to be a diagonal matrix with real non-negative elements 
by an $SU(N_f)$ similarity transformation preserving the form (\ref{Adiag}).
As a result, $M$ is of the form
\begin{equation}
  g^2M=\left(
    \begin{array}{cccccccc}
      \rho&      &      &    &\kappa_1&      &      &  \\
      &\ddots&      &    &        &\ddots&      &  \\
      &      &\ddots&    &        &      &\mms{\kappa_{N_f-N_c}}& \\
      &      &      &\rho&        &      &      &  \\
      &      &      &    &        &      &      & \\
      &      &      &    &        &      &      & \\
      &      &      &    &        &      &      &
    \end{array}
  \right),
  \label{Monbb}
\end{equation}
where $\kappa_i\in \R_+$ and $\rho\in \C$.

For the case $\Tr M=0$, namely $A=s=0$, we can diagonalize $Y$ by an $SU(N_f)$
similarity transformation:
\begin{equation}
  g^2M=\left(
    \begin{array}{cccc}
      \hs{5ex}&\kappa_1 \\
      &         &\ddots \\
      &         &      &\kappa_r \hs{5ex}  \\
      \\
      \\
      \\
    \end{array}\right),
  \label{Monnb}
\end{equation}
where $\kappa_i\in\R_+$ and $r\leq [N_f/2]$.

Note that these solutions imply a useful equation
$m\cdot M=0$.

Let us consider about the baryons.
For the case $\Tr M=0$, since $\rank M\leq [N_{f}/2]<N_c$, (\ref{BBM}) 
implies either $B=0$ or $\wt{B}=0$. Without loss of generality,
we can set $\wt{B}=0$
and the form of $M$ to be as in (\ref{Monnb}) with $\kappa_i\neq 0$.
Then (\ref{MB0}), (\ref{BM0}) and $m\cdot B=0$ imply that
the only non-zero elements of $B^{i_1\cdots i_{N_c}}$ are
$B^{12\cdots r i_{r+1}\cdots i_{N_c}}$ with
$2r<i_{r+1}<i_{r+2}\cdots<i_{N_c}\leq N_f$
(up to permutations of the flavor indices). So we find $B=0$ for $N_f-N_c<r$.
For $N_f-N_c\geq r$, $B$ can be non-zero. Using (\ref{BB0})
\footnote{This constraint is so called Pl\"ucker relation, and
  so we can apply Pl\"ucker embedding theorem.}
and a flavor symmetry,
which preserves the form of the meson (\ref{Monnb}),
we can reduce the baryon to only one element, say
$B^{1\cdots r,2r+1\cdots N_c+r}$. Using $U(1)_B$ symmetry, it can be chosen
to be real.
For the case $\Tr M\neq 0$,
the baryons are non-zero and
expressed by the mesons using (\ref{BBM}) up to the ratio
of $B$ and $\wt{B}$.

In summary, we have obtained two types of branches: the baryonic and
the non-baryonic branches.
\begin{itemize}
\item [$<1>$]{\bf The Baryonic Branch} \\
$B\neq 0$ or $\wt{B}\neq 0$
\footnote{From (\ref{SkB0}), $B\neq 0$ or $\wt{B}\neq 0$ implies
  $\Phi=0$.},
and the meson is as in (\ref{Monbb}).
We also include the limit $B,\wt{B}\ra 0$.
This branch exists for $N_c\leq N_f$.
\item [$<2>$] {\bf The non-Baryonic Branch} \\
$B=\wt{B}=0$ and the meson is as in (\ref{Monnb}).
\end{itemize}

As a check,
 we can determine the vacuum moduli space directly
using the F-term and D-term equations for $Q$, $\wt{Q}$ and $\Phi$, and 
get the same result as above (see appendix \ref{SU}). 

Next we investigate the magnetic theory and find a correspondence
of the gauge invariant operators between the electric and
magnetic theory.

The meson and the baryons in the magnetic theory are
\footnote{Note that our definition of the baryons ($b,\wt{b}$)
  are different from those in \cite{APS}. The baryons ($b,\wt{b}$)
  in \cite{APS} are defined in terms of the low
  energy effective theory at the root of the baryonic branch
  for $N_f<2N_c$.
  }
,
\begin{eqnarray}
  N_j^i&\equiv&q_{aj}\wt{q}^{ai}, \\
  b_{i_1\cdots i_{N_c}}&\equiv&q_{a_1i_1}\cdots q_{a_{N_c}i_{N_c}}
  \epsilon^{a_1\cdots a_{N_c}}, \\
  \wt{b}^{i_1\cdots i_{N_c}}&\equiv&
\wt{q}^{a_1i_1}\cdots \wt{q}^{a_{N_c}i_{N_c}}
  \epsilon_{a_1\cdots a_{N_c}}. 
\end{eqnarray}

The constraints for the meson and the baryons are
\begin{eqnarray}
  (*b)\wt{b}&=&*(N^{N_c}),
  \label{bbN} \\
  (*\wt{b})\cdot N&=&N\cdot *b=0,
  \label{Nb0} \\
  (*\wt{b})\cdot\wt{b}&=&b\cdot *b=0,
  \\
  N\cdot N'&=&0, 
  \\
  N'\cdot \wt{b}&=&b\cdot N'=0, 
  \\
  **(\wt{m}\cdot b)&=&*(\wt{m}\cdot \wt{b})=0,
  \\
  \wt{m}\cdot N&=&N\cdot\wt{m}
  \\
  \wt{m}^i_jN^j_i&=&0.
  \label{TrmN0}
\end{eqnarray}

We can see that the Higgs branches in the electric theory and those
in the magnetic theory are the same under the correspondence of the gauge
invariant operators:
\begin{eqnarray}
  \mbox{electric} & \lra & \mbox{magnetic}  \nn\\
  g^2M          & \lra &  \wt{g}^2N'           \\
  g^{N_c} B     & \lra &  (-\wt{g})^{N_c} * b \\
  g^{N_c}\wt{B} & \lra &  \wt{g}^{N_c} * \wt{b}.
\end{eqnarray}
We must check that the correspondence is compatible with all the
constraints.
It is easy to show that (\ref{BM0})$\sim$(\ref{TrmM})
imply (\ref{Nb0})$\sim$(\ref{TrmN0}). To see that (\ref{BBM}) implies
(\ref{bbN}), we use the solutions (\ref{Monbb}) and (\ref{Monnb}).
On the baryonic branch, for example, we have
\begin{equation}
  \wt{g}^2N\lra g^2M'=
  \left(
    \begin{array}{cccccccc}
      &   &   &    &\kappa_1&      &      & \\
      &   &   &    &        &\ddots&      & \\
      &   &   &    &        &      &\makebox[4ex]{$\kappa_{N_f-N_c}$}& \\
      &   &   &    & -\rho  &      &      & \\
      &   &   &    &        &\ddots&      & \\
      &   &   &    &        &      &\ddots& \\
      &   &   &    &        &      &      &-\rho
    \end{array}
  \right).
  \label{Nonbb}
\end{equation}
Now it is easy to see that
\begin{eqnarray}
  &&\wt{g}^{2N_c}N^{[i_1}_{j_1}\cdots N^{i_{N_c}]}_{j_{N_c}} \nn\\
  &\lra &
  g^{2N_c}M'^{[i_1}_{j_1}\cdots M'^{i_{N_c}]}_{j_{N_c}}=
  (-g^2)^{N_c}\epsilon^{i_1\cdots i_{2N_c}}\epsilon_{j_1\cdots j_{2N_c}}
  M^{j_{N_c+1}}_{i_{N_c+1}}\cdots M^{j_{2N_c}}_{i_{2N_c}}
\end{eqnarray}
and that (\ref{BBM}) implies (\ref{bbN}).
Thus we conclude that the Higgs branches
in the electric and magnetic theory are the same. 
 Combining with the fact that 
the Coulomb branches in the electric and magnetic theory are also the
same, as mentioned in section \ref{S-duality}, it means that
the vacuum moduli spaces are exactly the same. 
This is one of the most non-trivial evidences for the existence of
the S-duality.

\subsection{the baryonic root}
Before closing this section, we want to comment on the unbroken gauge group
at the baryonic root
\footnote{
The baryonic root is a point where $M=B=\wt{B}=0$ on the baryonic branch. 
}.
We now consider the baryonic branch with $\Tr M\neq 0$ in the electric
theory. 
{}From the F-term equations (\ref{F1})$\sim$(\ref{F3}), we have
\begin{equation}
  \left(
    \frac{1}{N_c}\Tr M
  \right)\tr(\Phi^k)=\sum_{i=1}^{2N_c}(-\frac{m_i}{g})^k M^i_i.
  \label{U}
\end{equation}
Recall that we have set $m=\diag(0,\cdots,0,m_{N_f+1},\cdots,m_{2N_c})$,
and $M$ can be written as in (\ref{Monbb}).
So,  the right hand side of (\ref{U}) vanishes, because $N_c\leq N_f$,
and (\ref{U}) implies 
$\tr(\Phi^k)=0$. Taking the limit $M,B,\wt{B}\ra 0$
along this branch,
we expect $SU(N_c)$ gauge symmetry unbroken.

On the other hand, similar equation in the magnetic theory is
\begin{equation}
  \left(\frac{1}{N_c}\Tr N\right)\tr(\varphi^k)=\sum_{i=1}^{2N_c}
  (-\frac{\wt{m}_i}{\wt{g}})^k N^i_i.
  \label{u}
\end{equation}
Substituting the form of $N$ on the baryonic branch (\ref{Nonbb}), we have
\begin{equation}
  \rho\tr(\varphi^k)=\rho\sum_{i=N_c+1}^{2N_c}(-\frac{\wt{m}_i}{\wt{g}})^k.
\end{equation}
For $\rho\neq 0$, we find
$\wt{g}^k\tr(\varphi^k)=\sum_{i=N_c+1}^{2N_c}(-\wt{m}_i)^k$,
which implies the form
\begin{equation}
  \wt{g}\varphi=\diag(2m_s,\cdots,2m_s,2m_s-m_{N_f+1},\cdots,2m_s-m_{2N_c})
  \label{varphi}
\end{equation}
up to permutations
\footnote{
  This form is exactly the same as that in \cite{APS} which is
  deduced by the requirement of IR-freedom and the existence of a purely
  hypermultiplet Higgs branch.
  }.
Now the unbroken gauge group $SU(N_f-N_c)\times U(1)^{2N_c-N_f}$
is expected at the
limit $b,\wt{b},N\ra 0$. This result is consistent with \cite{APS}.

\section{Deformations of ${\cal N}=2$ Theory to ${\cal N}=1$ Theory}

In this section, we will see how the electric and magnetic theory are
deformed when we 
break ${\cal N}=2$ supersymmetry to ${\cal N}=1$ adding the adjoint mass
term and how the duality changes.

\subsection{${\cal N}=1$ deformed electric theory}\label{electric}
We now add the adjoint mass term $\mu \tr \Phi^2$, which breaks ${\cal N}=2$
supersymmetry to ${\cal N}=1$ explicitly, to the superpotential (\ref{Wele})
\begin{equation}
  W_{\ele}=\sq gQ^i\Phi \wt{Q}_i +\sq m_iQ^i\wt{Q}_i
  +\frac{\mu}{\sqrt{2}}\tr \Phi^2.
  \label{WeleN1}
\end{equation}
The F-term equation (\ref{F1}) is modified
\begin{equation}
  g\left(Q^i_a\wt{Q}^b_i-\frac{1}{N_c}(Q^i\wt{Q}_i)\delta^b_a\right)
  +\mu\Phi^b_a=0.
  \label{PhibyQ}
\end{equation}
We can eliminate $\Phi$ using this equation. The
F-term equations (\ref{F2}) and (\ref{F3}) become
\begin{equation}
  \wt{Q}_i^a(\wt{M})^i_j=(\wt{M})^i_jQ^j_a=0,
\end{equation}
where we have defined $g^2\wt{M}\equiv g^2M'-\mu m$.
The constraints (\ref{MM0}) and (\ref{MB0}) are modified to be
\begin{eqnarray}
  M\cdot \wt{M}&=&0,  \label{MM''0}\\
  \wt{M}\cdot B&=&\wt{B}\cdot\wt{M}=0.   \label{M''B0}
\end{eqnarray}
Other constraints (\ref{BBM}), (\ref{BM0}), (\ref{BB0}),
(\ref{mB0}) and (\ref{mMMm})
 are not
modified, while (\ref{TrmM}) is modified to a redundant constraint. 

Since we have broken ${\cal N}=2$ supersymmetry, the Higgs branches will
receive quantum corrections. But we analyze classically for the time
being and later we consider the quantum effects.

Now we study the classical moduli space. As (\ref{BBM}),
(\ref{BM0}) and (\ref{mMMm}) are not modified,
$M$ can be put into the form (\ref{M1})
\begin{equation}
  g^2 M =
  \left(
    \begin{array}{cccccc}
      X & Y &     &        &     & \\
      0 & 0 &     &        &     & \\
      &   & \mms{d_{N_f+1}} &        &     & \\
      &   &     & \ddots &     & \\
      &   &     &        & \mms{d_{N_f+s}} & \\
      &   &     &        &     &0
    \end{array}
  \right).
\end{equation}
Then (\ref{MM''0}) implies that
\begin{eqnarray}
  X^i_j&=&\frac{1}{N_c}(\Tr X+\sum_{k=1}^s d_{N_f+k})\delta^i_j, 
  \\
  d_{N_f+i}&=&\frac{1}{N_c}(\Tr X+\sum_{k=1}^s d_{N_f+k})
  +\mu m_{N_f+i}.
\end{eqnarray}
These equations imply 
\begin{equation}
  (N_c-r-s)g^2\Tr M=N_c \mu(\sum^s_{i=1}m_{N_f+i}),
  \label{TrM}
\end{equation}
which has three classes of solutions.
\begin{itemize}
\item [$<1>$]{\bf The Baryonic Branch} ($r=N_c ,s=0$) \\
  For the case $r+s=N_c$, as we have chosen bare masses to be generic,
  (\ref{TrM}) implies that $s=0$ and $M$ is of the form
  \begin{equation}
    g^2M=\left(
      \begin{array}{cccccccc}
        \rho&      &      &    &\kappa_1&      &      &  \\
        &\ddots&      &    &        &\ddots&      &  \\
        &      &\ddots&    &        &      &\kappa_{N_f-N_c}& \\
        &      &      &\rho&        &      &      &  \\
        &      &      &    &        &      &      & \\
        &      &      &    &        &      &      & \\
        &      &      &    &        &      &      & \\
        &      &      &    &        &      &      &
      \end{array}
    \right),
    \label{MrNc}
  \end{equation}
  where $\rho\in\C$ and $\kappa_i\in\R_+$. \\
\item [$<2>$] {\bf The non-Baryonic Branch}
 ($r<N_c, s=0$) \\
  For the case $r+s< N_c$ and $s=0$, we get
  \begin{equation}
    g^2M=\left(
      \begin{array}{cccc}
        \hs{5ex}&\kappa_1 \\
        &         &\ddots \\
        &         &      &\kappa_r \hs{5ex}  \\
        \\
        \\
        \\
      \end{array}\right),
  \end{equation}
  where $\kappa_i\in\R_+$ and $r\leq[N_f/2]$. \\
\item [$<3>$] {\bf The Exceptional Branch} ($s\neq 0$) \\
  For the case $r+s<N_c$ and $s\neq 0$, the value of $\Tr M$ is fixed:
  \begin{equation}
    c\equiv\frac{g^2}{N_c}\Tr M=\frac{\mu}{N_c-r-s}\sum^s_{i=1}m_{N_f+i},
  \end{equation}
  and $M$ becomes
  \begin{equation}
    g^2M=
    \left(
      \begin{array}{ccccccccccccc}
        c  &      &&    &\kappa_1&      &        & &    &      &    &    &  \\
        &\ddots&&    &        &\ddots&        & &    &      &    &    &  \\
        &    &\ddots&&        &      &\kappa_{r'}&&  &      &    &    &  \\
        &      && c  &        &      &        & &    &      &    &    &  \\
        &      &&    &        &      &        & &    &      &    &    &  \\
        &      &&    &        &      &        & &    &      &    &    &  \\
        &      &&    &        &      &        & &    &      &    &    &  \\
        &      &&    &        &      &        & &d_{N_f+1}& &    &    &  \\
        &      &&    &        &      &        & &    &\ddots&    &    &  \\
        &      &&    &        &      &        & &    &      &d_{N_f+s}&& \\
        &      &&    &        &      &        & &    &      &    &    &  \\
        &      &&    &        &      &        & &    &      &    &    &  
      \end{array}
    \right),
  \end{equation}
  where $\kappa_i \in \R_+, d_{N_f+i}=c+\mu m_{N_f+i}$ and $r'\leq$ min
$\{N_f-r,r\}$.\\
\end{itemize}

The baryons take the same form as in the ${\cal N}=2$ case
on the baryonic and non-baryonic branches, while these are all zero 
on the exceptional branch.
We can show that these solutions are the same as the solutions derived
from the F-term and
D-term equations (see appendix \ref{SU2}), and that the modified
constraints, which we have considered, form a complete set of constraints.

We then consider non-perturbative effects.
On the baryonic branch the gauge group is broken completely
and the theory is weakly coupled. So we expect that this branch will
not be lifted quantum mechanically.
On the non-baryonic branch of $\rank M = r$ the gauge group is broken to
$SU(N_c-r)$ and the number of the massless quarks are
$N_f-2r$
\footnote{
In order to show these facts, it may be useful to see the explicit
form of $Q$ and $\wt{Q}$, which is listed in appendix \ref{SU2}.
}.
For $r>N_f-N_c$, Affleck-Dine-Seiberg superpotential
is generated \cite{ADS}, and the classical vacua are lifted.
The non-baryonic
branch of  $r\leq N_f-N_c$ is a submanifold
 of the baryonic branch, {\it i.e.} $\rho=0$ in (\ref{MrNc}).
On the exceptional branch the theory become ${\cal N}=1$ $SU(N_c-r-s)$
Super Yang-Mills theory with several massless singlets. Then owing to
the gaugino condensation, the dynamical superpotential is generated
and we expect that this branch will disappear quantum mechanically.

As a result, we conclude that only the baryonic branch remains as 
the vacuum moduli space.

\subsection{${\cal N}=1$ deformed magnetic theory and duality}\label{magnetic}

Before presenting the answer, it would be
instructive to show how we find the duality. Notice that the equations
(\ref{MM''0}), (\ref{BM0}) and (\ref{M''B0}) are symmetric under
Hodge dual transformations for the baryons ($B\ra *B, \wt{B}\ra
*\wt{B}$) and interchanging $M$ and $\wt{M}$. This fact suggests that
there is a dual (magnetic) theory in which the meson $\wt{g}^2N$ corresponds
to
$g^2\wt{M}(=g^2M'-\mu m)$. Then we find that (\ref{MM''0}) implies
\begin{equation}
  \wt{N}\cdot N=0, \label{N''N0}
\end{equation}
where $\wt{g}^2\wt{N}\equiv \wt{g}^2N'+\mu \wt{m}$.
The superpotential, whose F-term equation implies (\ref{N''N0}), is
\begin{eqnarray}
  W_{\mag}=\sq \wt{g}q_i\varphi \wt{q}^i
  +\sq\wt{m}_iq_i\wt{q}^i-\frac{\mu}{\sq}\tr \varphi^2.
  \label{WmagN1}
\end{eqnarray}
Note that the sign of the adjoint mass term is different from that in
the electric theory. 

We claim that this theory is dual to the electric theory (\ref{WeleN1}).
The correspondence of the meson and the baryons is
\begin{eqnarray}
  \mbox{electric} & \lra & \mbox{magnetic} \nn       \\
  g^2M          & \lra &  \wt{g}^2\wt{N}
  \label{MbyN}  \\
  g^{N_c} B     & \lra &  (-\wt{g})^{N_c} * b \\
  g^{N_c}\wt{B} & \lra &  \wt{g}^{N_c} * \wt{b}.
\end{eqnarray}
This correspondence smoothly flows to that in  ${\cal N}=2$ theory
in the limit  $\mu \ra 0$.

The inverse map is
\begin{eqnarray}
\mbox{electric} & \lra & \mbox{magnetic} \nn    \\
  g^2\wt{M}     & \lra &  \wt{g}^2N        
\label{NbyM}\\
  g^{N_c}* B     & \lra &  \wt{g}^{N_c} b \\
 (-g)^{N_c}*\wt{B} & \lra &  \wt{g}^{N_c}\wt{b}.
\end{eqnarray}

The following correspondence of the constraints is almost trivial.
\begin{eqnarray}
  M\cdot\wt{M}=0      &\lra&  \wt{N}\cdot N=0 \\
  M\cdot(*B)=M\cdot(*\wt{B})=0      &\lra&  \wt{N}\cdot
  b=\wt{N}\cdot\wt{b}=0 \\
  \wt{M}\cdot B=\wt{M}\cdot\wt{B}=0 &\lra&
  N\cdot(*b)=N\cdot(*\wt{b})=0 \\
  m\cdot M=M\cdot m &\lra& \wt{m}\cdot N=N\cdot \wt{m}.
\end{eqnarray}
Now, we must check
\begin{equation}
  (*B)\wt{B}=*(M^{N_c}) \lra (*b)\wt{b}=*(N^{N_c}). \label{BB-bb}
\end{equation}
$M$ is of the form (\ref{MrNc}) and then $N$ is
\begin{eqnarray}
  \wt{g}^2N\lra g^2\wt{M}=
  \left(
    \begin{array}{cccccccccccccc}
      \hs{5ex}&\kappa_1&      &                \\
      &        &\ddots&                \\
      &        &      &\mms{\kappa_{N_f-N_c}}\\
      &-\rho   &      &                \\
      &        &\ddots&                \\
      &        &      &-\rho           \\
      &        &      &   &-\sigma_{N_f+1} \\
      &        &      &   &         &\ddots   \\
      &        &      &   &         &      &-\sigma_{2N_c}
    \end{array}
  \right),
\end{eqnarray}
where $\sigma_{N_f+i}=\rho+\mu m_{N_f+i}$.

Then we have
\begin{eqnarray}
  &&\hspace*{-8ex}\wt{g}^{2N_c}N^{[i_1}_{j_1}\cdots N^{i_{N_c}]}_{j_{N_c}}
\nn\\
  &\lra& g^{2N_c}\wt{M}^{[i_1}_{j_1}\cdots
  \wt{M}^{i_{N_c}]}_{j_{N_c}}\nn\\
 &&=
  (-g^2)^{N_c}
  \left(
    \prod^{2N_c}_{i=N_f+1}\frac{\sigma_i}{\rho}
  \right)
  \epsilon^{i_1\cdots i_{2N_c}}\epsilon_{j_1\cdots j_{2N_c}}
  M^{j_{N_c+1}}_{i_{N_c+1}}\cdots M^{j_{2N_c}}_{i_{2N_c}}.
\end{eqnarray}

So, if we choose the normalization factor for baryons as
\begin{eqnarray}
  b_{i_1\cdots i_{N_c}} &=&
  \left(\prod_{i=N_f+1}^{2N_c}\frac{\rho}{\sigma_i}\right)^{1/2}
  \epsilon_{a_1,\ldots ,a_{N_c}}q_{i_1}^{a1}\cdots q_{i_{N_c}}^{a_{N_c}}, \\
  \wt{b}^{i_1\cdots
    i_{N_c}}&=&\left(\prod_{i=N_f+1}^{2N_c}\frac{\rho}{\sigma_i}\right)^{1/2}
  \epsilon^{a_1,\ldots ,a_{N_c}}\wt{q}^{i_1}_{a1}\cdots
  \wt{q}^{i_{N_c}}_{a_{N_c} },
\end{eqnarray}
\footnote{There is a subtlety for the case $\rho=0$ or $\sigma_i=0$, but  
this arises only for submanifolds of the baryonic branch. 
  }
where 
\begin{eqnarray}
  \rho&=& -\frac{\wt{g}^2}{N_c}\Tr N-2\mu m_S
  \lra\frac{g^2}{N_c}\Tr M, \\
  \sigma_i&=&-\frac{\wt{g}^2}{N_c}\Tr N+\mu \wt{m}_i
  \lra \frac{g^2}{N_c}\Tr M+\mu m_i,
\end{eqnarray}
we get the correspondence (\ref{BB-bb}).
Note that the normalization factor can be rewritten in the flavor singlet
form:
\begin{equation}
  \prod_{i=N_f+1}^{2N_c}\frac{\sigma_i}{\rho}
  =\prod_{i=1}^{2N_c}\frac{\sigma_i}{\rho}
  =\frac{\det (\sigma^i_j)}{\rho^{2N_c}},
\end{equation}
where we have defined
\begin{equation}
  \sigma^i_j\equiv-\frac{\wt{g}^2}{N_c}\Tr N\delta^i_j+\mu \wt{m}^i_j.
\end{equation}

If there were the non-baryonic branches of $r\geq N_f-N_c$ or the
exceptional branches in the electric theory,
the rank of the corresponding $N$ would become larger than
$N_c$ contradicting with (\ref{bbN}) and (\ref{Nb0}).

\subsection{the gauge group in the magnetic theory}
\label{SU(f-c)}

In the last subsection, we have seen that the vacuum moduli space
in the electric and magnetic theory are the same, and claimed that
there is a duality in the 
${\cal N}=1$ deformed theories.
Let us show that the magnetic theory
is ${\cal N}=1$ supersymmetric $SU(N_f-N_c)$ QCD as expected
from the ${\cal N}=1$ duality of N.Seiberg.

The F-term equations in the ${\cal N}=1$ deformed magnetic theory are
\begin{eqnarray}
  \wt{g}\left(q_{ai}\wt{q}^{bi}-\frac{1}{N_c}(q_i\wt{q}^i)\delta^b_a\right)
  &=& \mu\varphi^b_a , \label{f1} \\
  \wt{g} \varphi^a_b\wt{q}^{bj}+\wt{m}_i^j \wt{q}^{ai}&=&0,
  \label{f2}  \\
  \wt{g} q_{ai}\varphi^a_b +\wt{m}_i^jq_{bj}&=&0. \label{f3}
\end{eqnarray}
{}From these equations we have recursive relations of 
$u_{k}\equiv\tr\varphi^k$:
\begin{equation}
 u_{k}=\frac{\wt{g}^2}{\mu}
 \left(
   -\frac{1}{N_c}(\Tr N)u_{k-1}+\sum_i(-\frac{\wt{m_i}}{\wt{g}})^{k-1}N^i_i  
 \right) \label{u_k}
\end{equation}
with the initial condition $u_1=0$.
We are interested in the baryonic root where $SU(N_c)$
gauge symmetry is expected to be unbroken in the electric theory.
 The point $M=0$ corresponds to the point
$\wt{g}^2 N=-\mu m$ (see (\ref{NbyM})) in the magnetic theory.
At this point, the solution of (\ref{u_k}) is again
\begin{equation}
  u_k=\tr\varphi^k=\sum^{2N_c}_{i=N_c+1}(-\frac{\wt{m}_i}{\wt{g}})^k,
  \label{VEVu}
\end{equation}
and $\varphi$ is the same as in (\ref{varphi}).
So, the gauge group is broken by $\VEV{\varphi}$ to
$SU(N_f-N_c)\times U(1)^{2N_c-N_f}$, and the condensations of dual
quarks 
\begin{equation}
  \VEV{q_i\wt{q}^i}=-\frac{\mu m_i}{\wt{g}^2},~~(i=N_f+1,\cdots,2N_c)
  \label{VEVq}
\end{equation}
break
the $U(1)^{2N_c-N_f}$ factor.
As a result, the magnetic theory is ${\cal N}=1$ supersymmetric
$SU(N_f-N_c)$ QCD with $N_f$ flavors. 

\subsection{Leigh-Strassler transformation}
\label{LStr}
The magnetic theory of N.Seiberg's duality in ${\cal N}=1$
supersymmetric QCD is ${\cal N}=1$ supersymmetric
$SU(N_f-N_c)$ theory with $N_f$ flavors of dual quarks $q_i$, $\wt{q}^i$,
$(i=1,\cdots,N_f)$ and a gauge singlet meson field ${\cal M}^i_j$,
interacting with the dual quarks by the superpotential:
\begin{equation}
  W_{\mag}^{{\cal N}=1}=q_i{\cal M}^i_j\wt{q}^j.
  \label{Wmagmeson}
\end{equation}

So far, we have needed
neither the meson field ${\cal M}^i_j$
nor the superpotential $W_{\mag}^{{\cal N}=1}$.
So our derivation of ${\cal N}=1$
duality may seem to be inconsistent. But, it is not the case.
There was an argument given by
R.G.Leigh and M.J.Strassler \cite{LS}
which showed the origin of the meson field and the superpotential
for $N_f=2N_c$ case. We will now show that applying their argument
to the case with quark mass terms, there appears
the superpotential (\ref{Wmagmeson}) with the meson field ${\cal M}$.

The electric theory discussed in section \ref{electric} has the superpotential
(\ref{WeleN1}):
\begin{equation}
  W_{\ele}=\sq gQ^i\Phi\wt{Q}_i+\sq m_iQ^i\wt{Q}_i
  +\frac{\mu}{\sq}\tr\Phi^2.
\label{WeleN1'}
\end{equation}
Integrating out $\Phi$, ({\it i.e.} inserting (\ref{PhibyQ})), we get
\begin{equation}
  W_{\ele}
  =-\frac{g^2}{\sq\mu}\left( (Q^i\wt{Q}_j)( Q^j\wt{Q}_i)
    -\frac{1}{N_c}(Q^i\wt{Q}_i)^2\right)+\sq m_iQ^i\wt{Q}_i. 
\label{WeleQQ}
\end{equation}
We consider the theory around $Q^i=\wt{Q}_i=0$.
In the limit $g\ra 0$, we can neglect the first term:
\begin{equation}
  W_{\ele}\sim\sq m_iQ^i\wt{Q}_i. 
\end{equation}
Integrating out the massive quarks, the theory turns out to be
${\cal N}=1$ supersymmetric QCD with $N_f$ massless flavors.

On the other hand, the magnetic theory discussed in section \ref{magnetic}
has the superpotential (\ref{WmagN1}):
\begin{equation}
  W_{\mag}=\sq\wt{g}q_i\varphi\wt{q}^i+\sq\wt{m}_iq_i\wt{q}^i
  -\frac{\mu}{\sq}\tr\varphi^2.
\label{WmagN1'}
\end{equation}
Similarly, we integrate out $\varphi$, ({\it i.e.} inserting (\ref{f1})):
\begin{eqnarray}
  W_{\mag}
  &=&\frac{\wt{g}^2}{\sq\mu}\left( (q_i\wt{q}^j)( q_j\wt{q}^i)
    -\frac{1}{N_c}(q_i\wt{q}^i)^2\right)+\sq\wt{m}_iq_i\wt{q}^i  \\
  &=&q_i{\cal N'}^i_j\wt{q}^j+\sq \wt{m}_i(q_i\wt{q}^i)
  -\frac{\mu}{2\sq\wt{g}^2}
  \left(\Tr {\cal N'}^2-\frac{1}{N_c}(\Tr {\cal N'})^2\right),
\end{eqnarray}
where we have introduced an auxiliary field ${\cal N'}^i_j$, which can
be eliminated by the equation of motion (F-term equation)
\begin{equation}
  {\cal N'}^i_j=\frac{\sq\wt{g}^2}{\mu}
  \left( 
    q_j\wt{q}^i-\frac{1}{N_c}\Tr (q\wt{q})\delta^i_j
  \right)=\frac{\sq\wt{g}^2}{\mu}(N')^i_j.
\end{equation}
{}From the correspondence of the meson (\ref{MbyN}), it is clear that
we should rewrite the superpotential using
\begin{eqnarray}
  {\cal M}^i_j&\equiv&{\cal N'}^i_j+\sq \wt{m}^i_j \\
  &=&\frac{\sq}{\mu}\left( \wt{g}^2(N')^i_j+\mu \wt{m}^i_j\right)
  ~~\mbox{(on shell)}\\
  &\lra&\frac{\sq}{\mu}g^2M^i_j.
\end{eqnarray}
Then the superpotential becomes
\begin{equation}
  W_{\mag}=q_i{\cal M}^i_j\wt{q}^j+\frac{\mu m_i}{\wt{g}^2}{\cal M}^i_i
  -\frac{\mu}{2\sq\wt{g}^2}
  \left(\Tr {\cal M}^2-\frac{1}{N_c}(\Tr {\cal M})^2\right)
  +\mbox{const}.
\label{Wmagq}
\end{equation}
We take the same limit as above $\wt{g}\sim 1/g\ra \infty$,
fixing (\ref{VEVu}) and (\ref{VEVq}).
Then we can neglect the third term:
\begin{equation}
  W_{\mag}\sim q_i{\cal M}^i_j\wt{q}^j+\frac{\mu m_i}{\wt{g}^2}{\cal M}^i_i.
  \label{WmagmM}
\end{equation}
The equations of motion,
\begin{equation}
  q_i\wt{q}^i= -\frac{\mu m_i}{\wt{g}^2}\neq 0~~~(i=N_f+1,\cdots,2N_c)
\end{equation}
imply that the gauge group is broken to $SU(N_f-N_c)$, and the first term
in (\ref{WmagmM}) gives masses to $q_i,\wt{q}^j$, and ${\cal M}^i_j$ for
$(i,j=N_f+1,\cdots,2N_c)$. 
Thus the massless fields are
$N_f$ dual-quarks ($q_i,\wt{q}^i$) and
the meson field (${\cal M}^i_j$, ($i,j=1,\cdots,N_f$)) 
which are interacting with each other by the superpotential
(\ref{Wmagmeson}).
So, the theory is exactly the magnetic theory of N.Seiberg's duality.
Note that although the meson field ${\cal M}^i_j$ appeared as
an auxiliary field, we expect that this field becomes dynamical
as a result of quantum effects, otherwise 't Hooft anomaly matching
conditions are not satisfied \cite{S}.

Conversely, we can begin our argument with N.Seiberg's ${\cal N}=1$
duality.
Consider ${\cal N}=1$ supersymmetric $SU(N_c)$ QCD with $2N_c$
flavors and its dual. The meson field ${\cal M}$ in the magnetic
theory corresponds to $Q\wt{Q}$ in the electric theory. 
When we deform the superpotential in the electric theory as 
(\ref{WeleQQ}), 
the corresponding deformation of the superpotential in the magnetic theory
is as (\ref{Wmagq}). Introducing the adjoint auxiliary field, we can
rewrite the superpotential as (\ref{WeleN1'}) or (\ref{WmagN1'}).

\section{Duality in $SO(N_c)$ Gauge Theory}\label{SO}

\subsection{S-duality in ${\cal N}=2$ theory}
We consider in this subsection ${\cal N}=2$ supersymmetric $SO(N_c)$
QCD with $N_c-2$ hypermultiplets in the vector representation of
the gauge group.
We have chosen the number of the hypermultiplets so that the theory
is scale invariant, and has a dual description.
As in the $SU(N_c)$ case, we describe
the theory in terms of ${\cal N}=1$ superfields
\footnote
{We use the notations in \cite{APSh,APSh1}}
by a field strength chiral multiplet $W^{\alpha}_{ab}$ and a chiral multiplet
$\Phi_{ab}$, both in the adjoint representation of the gauge group,
and chiral multiplets $Q^i_a$ in the vector representation of the gauge group,
where $a,b=1,\cdots,N_c$ are color indices, and $i=1,\cdots,2(N_c-2)$ are
flavor indices. The superpotential is
\begin{equation}
  W_{\ele}=\sq gQ_a^i\Phi_{ab}Q_b^j\J_{ij}+\sq m_{ij}Q^i_aQ^j_a,
\end{equation}
where $\J\equiv ({0\atop -1}{1\atop 0})\otimes\I$ is the symplectic metric
and $\I$ is the $(N_c-2)\times(N_c-2)$ identity matrix.
We raise and lower the flavor indices by contracting with $\J$.
See appendix \ref{conv} for our conventions.

Since the pairs $(Q^i,Q^{N_c-2+i})$ make up ${\cal N}=2$ hypermultiplets,
the bare mass matrix  $m_{ij}$ is 
$m=(m_{ij})\equiv({0\atop 1}{1\atop 0})\otimes\diag(m_1,\cdots,m_{N_c-2})$.

As commented in section \ref{S-duality},
S-dualities in ${\cal N}=2$ supersymmetric $SO(N_c)$ theories are also known
\cite{APSh1,HO}. The hyperelliptic curve which describe the coulomb phase
of the theory is
\begin{equation}
y^2=x\prod_{a=1}^{[N_c/2]}(x-\phi_a^2)^2+4fx^{3-\epsilon}
\prod_{j=1}^{N_c-2}(x-m_j^2),
\end{equation}
where $f(\tau)=\theta_2^4\theta_4^4/(\theta_2^4-\theta_4^4)^2$ and
$\epsilon = (N_c \mod 2)$ \cite{APSh1,HO}.
This curve is invariant under $\tau\ra -1/\tau$, and so we expect
that the magnetic theory is also ${\cal N}=2$ supersymmetric $SO(N_c)$ QCD
with $N_c-2$ hypermultiplets in the vector representation of the gauge group,
and has the superpotential
\begin{equation}
W_{\mag}=\sq \wt{g}q^a_i\varphi^{ab}q^b_j\J^{ij}+\sq m^{ij}q_i^aq_j^a,
\end{equation}
where $\wt{g}=1/g$.

\subsection{${\cal N}$=1 deformed electric theory}
As in the $SU$ case, we add the adjoint mass term and break
${\cal N}=2$ supersymmetry to ${\cal N}=1$ explicitly:
\begin{equation}
  W_{\ele}=\sqrt{2}gQ_a^i\Phi_{ab}Q_b^j\J_{ij}+\sqrt{2}m_{ij}Q^i_aQ^j_a
  +\frac{\mu}{\sqrt{2}}\tr\Phi^2.
  \label{WeleSO}
\end{equation}

The F-term equations are
\begin{eqnarray}
  g Q^i_a\J_{ij}Q^j_b-\mu\Phi_{ab}&=&0,
  \label{SOOF1} \\
  g \Phi_{ab}Q^j_b\J_{ij}+m_{ij}Q^j_a&=&0.
  \label{SOOF2}
\end{eqnarray}
{}From (\ref{SOOF1}), we can eliminate $\Phi$, and so the vacuum moduli space
is
parameterized by the meson and the baryon which are defined by
\begin{eqnarray}
  M^{ij}&\equiv&Q^i_aQ^j_a,\\
  B^{i_1\cdots i_{N_c}} &\equiv&
  Q^{i_1}_{a_1}\cdots Q^{i_{N_c}}_{a_{N_c}}\epsilon^{a_1\cdots a_{N_c}}.
\end{eqnarray}
The constraints of these operators following from the definitions
are
\begin{eqnarray}
  (*B)B&=&*(M^{N_c}),
  \label{B0B0M} \\
  M\cdot *B&=&0,
  \label{MB00} \\
  (*B)\cdot B&=&0.
\end{eqnarray}
Here the meaning of ``${*}$'' and ``$\cdot$'' are 
almost the same as that
defined in section \ref{N2duality},
but for $*(M^{N_c})$ we mean 
$(\epsilon_{i_1\cdots i_{2N_c-4}}M^{i_1j_1}\cdots M^{i_{N_c}j_{N_c}})$.

Inserting (\ref{SOOF1}) into (\ref{SOOF2}), we get
\begin{equation}
  Q^i_a\J_{ij}\wt{M}^{jk}=0,
\end{equation}
where we have defined
$g^2\wt{M}^{jk}\equiv g^2M^{jk}-\mu m^{jk}$
and $m^{jk}\equiv\J^{jj'}m_{j'k'}\J^{k'k}(=m_{jk})$.
The constraints following from this equation are
\begin{eqnarray}
  M^{ij}\J_{jk}\wt{M}^{kl}&=&0,
  \label{MJM0} \\
  \wt{M}\cdot\J\cdot B&=&0,
   \\
  **(m\cdot \J\cdot B)&=&0.
\end{eqnarray}

(\ref{B0B0M}) and (\ref{MB00}) imply $\rank M\leq N_c$.
{}From (\ref{B0B0M}) it follows that $B$ is zero for $\rank M< N_c$
and $B$ can be expressed by $M$ for $\rank M=N_c$.
It is easy to check that the other constraints including $B$ are redundant.
Thus the vacuum moduli space is parameterized by $M$ only, and the
constraints are (\ref{MJM0}) and $\rank M\leq N_c$.

We can determine the vacuum moduli space.
(\ref{MJM0}) implies
\begin{equation}
  M^{ij}\J_{jk}m^{kl}=m^{ij}\J_{jk}M^{kl}.
  \label{MJmmJM}
\end{equation}
We set 
$m^{ij}=\left({0\atop 1}{1\atop 0}\right)
\otimes\diag(0,\cdots,0,m_{N_f+1},\cdots,m_{N_c-2})$, where $m_i's$
are chosen to be generic, and then (\ref{MJmmJM})implies
\begin{equation}
  g^2M=\left(
    \begin{array}{ccccc}
      X  & 0  &\vline& Y  & 0  \\
      0  & 0  &\vline& 0  & d  \\
      \hline
      Y^T & 0  &\vline& Z  & 0  \\
      0  & d  &\vline& 0  & 0
    \end{array}
  \right),
\end{equation}
where $X$ and $Z$ are $N_f\times N_f$ symmetric matrices,
$Y$ is an $N_f\times N_f$ matrix
and $d=\diag(d_{N_f+1},\cdots,d_{N_c-2})$.
{}From (\ref{MJM0}) we find
\begin{eqnarray}
  \left(
    {X \atop Y^T}{Y \atop Z}
  \right)
  \left(
    { 0 \atop -\I}{\I \atop 0}
  \right)
  \left(
    {X \atop Y^T}{Y \atop Z}
  \right)
  &=&0,  \label{ABCD}  \\
  d_i(d_i-\mu m_i)&=&0,~~~i=N_f+1,\cdots,N_c-2.
\end{eqnarray}
It was shown in \cite{APSh} that the solutions of (\ref{ABCD})
can be reduced to $X=\diag(a_1,\cdots,a_{N_f})$, $(a_i\in\R_+)$, $Y=Z=0$,
using
similarity transformations of massless flavor symmetry group $USp(2N_f)$.
So we get
\begin{equation}
  g^2M=\left(
    \begin{array}{ccccc}
      X  & 0  &\vline& 0  & 0  \\
      0  & 0  &\vline& 0  & d  \\
      \hline
      0  & 0  &\vline& 0  & 0  \\
      0  & d  &\vline& 0  & 0
    \end{array}
  \right),
  \label{s-branch}
\end{equation}
where $X=\diag(a_1,\cdots,a_r,0,\cdots,0)$, $a_i\in\R_+$,
$d=\diag(\mu m_{N_f+1},\cdots,\mu m_{N_f+s},0,\cdots,0)$ and $r+2s\leq N_c$.

This form is the same as that derived from the F-term and D-term equations
for $Q$ and $\Phi$
(see appendix (\ref{so})),and so the constraints form a complete set of
the classical constraints. 

Let us consider the quantum effects. When the meson is as
(\ref{s-branch}), the theory is ${\cal N}=1$ supersymmetric
$SO(N_c-r-2s)$ QCD with $2(N_f-r)$
massless quarks in the vector representation. If $2(N_f-r)\leq
(N_c-r-2s)-5~~(i.e.~r-2s>2N_f-N_c+4)$, Affleck-Dine-Seiberg type
dynamical superpotential is generated and the classical vacua are lifted
\cite{IS}.
Therefore the vacuum moduli space consists of the branches of $r-2s\leq 
2N_f-N_c+4$.

The structure of the vacuum moduli space is rather different from that 
in the $SU(N_c)$ theory. 
 We have seen in section \ref{electric} that
there is essentially one branch (baryonic branch) in the $SU(N_c)$ theory.
In the $SO(N_c)$ theory, however, 
we have found several distinct branches
\footnote{Here we call a connected component of the vacuum moduli
  space by a branch.}, for
$d$ in (\ref{s-branch}) is a fixed matrix. Each branch has a point
with unbroken $SO(N_c-2s)$ gauge symmetry ($X=0$ in (\ref{s-branch})).
So we have a series of $SO$ theories ($i.e.$
$SO(N_c),SO(N_c-2),\cdots,SO(2N_f-N_c+4)$) all at once.

\subsection{${\cal N}$=1 deformed magnetic theory and duality}
The superpotential of the ${\cal N}=1$ deformed magnetic theory is
\begin{eqnarray}
W_{\mag}&=&\sq\wt{g}q^a_i\varphi^{ab}q^b_j\J^{ij}+\sq m^{ij}q^a_iq^a_j
-\frac{\mu}{\sq}\tr\varphi^2,
\end{eqnarray}
where $\wt{g}\equiv 1/g$.
We define the meson and the baryon as follows.
\begin{eqnarray}
N_{ij}&\equiv&q^a_iq^a_j,\\
b_{i_1\cdots i_{N_c}} &\equiv&
q^{a_1}_{i_1}\cdots q^{a_{N_c}}_{i_{N_c}}\epsilon^{a_1\cdots a_{N_c}}.
\end{eqnarray}
As in the electric theory, the baryon is redundant.

We claim that the magnetic theory is dual to the electric theory under
the correspondence:
\begin{eqnarray}
\mbox{electric}&\lra&\mbox{magnetic} \nn \\
g^2M^{ij}&\lra&\wt{g}^2\wt{N}^{ij}
\equiv\wt{g}^2\J^{ii'}\wt{N}_{i'j'}\J^{j'j} \\
(g^2\wt{M}^{ij}&\lra&\wt{g}^2N^{ij}),\nn
\end{eqnarray}
where $\wt{g}^2\wt{N}_{ij}\equiv \wt{g}^2N_{ij}+\mu m_{ij}$.

It is trivial to check the correspondence of the constraint:
\begin{eqnarray}
M^{ij}\J_{jk}\wt{M}^{kl}=0&\lra&\wt{N}_{ij}\J^{jk}N_{kl}=0.
\end{eqnarray}
When the meson is as (\ref{s-branch}), the rank of $N$ is
$r+2(N_c-2-N_f-s)$. As explained in the last subsection, we know that
$r-2s\leq 2N_f-N_c+4$, which implies $\rank N\leq N_c$.
As a result, we conclude that the vacuum moduli space in the electric
and magnetic theory are the same.

\subsection{the correspondence of the gauge group}
\label{SOgauge}

We can determine the gauge groups
in the electric and magnetic theory, and show that 
the duality in the last subsection is
consistent with N.Seiberg's duality.

The F-term equation (\ref{SOOF1}) implies
\begin{equation}
  \tr \Phi^k=\left(\frac{g}{\mu}\right)^k\Tr(\J\cdot M)^k.
\end{equation}
If we choose the branch in (\ref{s-branch}), and take $X=0$ then
\begin{equation}
  g^k\tr \Phi^k=\left\{\begin{array}{lc}
      0&(k:\mbox{ odd})\\
      2\sum_{i=1}^{s}(m_{N_f+i})^k
      &(k:\mbox{ even}).
    \end{array}\right.
\end{equation}
{}From this, we have
\begin{equation}
  g\Phi=
  \left(\begin{array}{ccccccccccc}
      0&      & &          &         &      &          &         & & &\\
      &\ddots& &          &         &      &          &         & & &\\
      &      &0&          &         &      &          &         & & &\\
      &      & &          &\mms{im_{N_f+1}}&      &          &         & & &\\
      &      & &\mms{-im_{N_f+1}}&         &      &          &         & & &\\
      &      & &          &         &\ddots&          &         & & &\\
      &      & &          &         &      &          &\mms{im_{N_f+s}}& & &\\
      &      & &          &         &      &\mms{-im_{N_f+s}}&         & & &\\
      &      & &          &         &      &          &         &0& &\\
      &      & &          &         &      &          &         & &\ddots&
    \end{array}\right).
\end{equation}
So we expect $SO(N_c-2s)$ gauge symmetry unbroken
\footnote{$U(1)$ factors are broken by $M\neq0$ (see section
  \ref{SU(f-c)}).}, and there are $2N_f$ massless quarks in
the vector representation of the unbroken gauge group.

On the other hand, on the corresponding branch in the magnetic theory,
\begin{equation}
  \wt{g}^k\tr
  \varphi^k=\left(-\frac{\wt{g}^2}{\mu}\right)^k\Tr(\J^{-1}\cdot N)^k 
  =\left\{\begin{array}{lc}
      0&(k:\mbox{ odd})\\
      2\sum_{i=N_f+s+1}^{N_c-2}(m_{i})^k
      &(k:\mbox{ even}),
    \end{array}\right.
\end{equation}
implying
\begin{equation}
  \wt{g}\varphi=
  \left(\begin{array}{ccccccccc}
     0&      &      & &            &           &      &          &          \\
      &\ddots&      & &            &           &      &          &          \\
      &      &\ddots& &            &           &      &          &          \\
      &      &      &0&            &           &      &          &          \\
      &      &      & &            &\mms{im_{N_f+s+1}}&   &      &          \\
      &      &      & &\mms{-im_{N_f+s+1}}&    &      &          &          \\
      &      &      & &            &           &\ddots&          &          \\
      &      &      & &            &           &      &    &\mms{im_{N_c-2}}\\
      &      &      & &            &           &      &\mms{-im_{N_c-2}}&   \\
    \end{array}\right).
\end{equation}
Thus the magnetic theory is expected to be $SO(2N_f-(N_c-2s)+4)$
 gauge theory with $2N_f$ massless
dual quarks in the vector representation.
This result supports N.Seiberg's duality in ${\cal N}=1$ supersymmetric
$SO$ QCD
\footnote{Here $N_f$ is the number of the massless hypermultiplets,
  and so the number of ${\cal N}=1$ chiral multiplets in the
vector representation of the gauge group is $2N_f$.}
\cite{S,IS,APSh}. 

Note that unlike the $SU$ case,
 in which we have obtained a duality for $SU(N_c)\lra SU(N_f-N_c)$,
we have obtained a series of the duality for $SO(N_c-2s)\lra
SO(2N_f-(N_c-2s)+4)$, ($s=0,1,\cdots, N_c-2-N_f$).

\subsection{Leigh-Strassler transformation}
Similar arguments as in section \ref{LStr} can be applied to the
$SO(N_c)$ theory, and we can show that the meson 
field and the superpotential are needed in the magnetic theory.
For simplicity, we investigate the duality on the branch of $s=0$ in
(\ref{s-branch}).

Integrating out $\Phi$ in the electric theory (\ref{WeleSO}), we have
\begin{equation}
  W_{\ele}=\frac{g^2}{\sq\mu}(Q^iQ^j)\J_{ik}\J_{jl}(Q^kQ^l)+\sq m_{ij}Q^iQ^j.
\end{equation}
If we take the limit $g\ra 0$, we can neglect
the first term:
\begin{equation}
  W_{\ele}\sim \sq m_{ij}Q^iQ^j.
\end{equation}
Integrating out massive flavors, the theory turns out to be ${\cal N}=1$
supersymmetric $SO(N_c)$ QCD with
$2N_f$ massless chiral multiplets in the vector representation of the gauge
group.

In the magnetic theory, on the other hand, integrating out $\varphi$, we get
\begin{equation}
W_{\mag}=-\frac{\wt{g}^2}{\sq\mu}(q_iq_j)\J^{ik}\J^{jl}(q_kq_l)
+\sq m^{ij}q_iq_j.
\end{equation}
We introduce an auxiliary field ${\cal M}^{ij}$ and
rewrite this superpotential:
\begin{equation}
W_{\mag}={\cal M}^{ij}q_iq_j+\frac{\mu}{\wt{g}^2} m_{ij}{\cal M}^{ij}
         -\frac{\mu}{2\sq\wt{g}^2}{\cal M}_{ij}{\cal M}^{ij}.
\end{equation}
The equation of motion implies
$\mu{\cal M}^{ij}=\sq(\wt{g}^2N^{ij}+\mu m^{ij})\lra\sq g^2M^{ij}$.
In the limit $\wt{g}\sim 1/g\ra\infty$, fixing $\VEV{qq}\sim \mu
m/\wt{g}^2$ and $\VEV{\varphi}\sim m/\wt{g}$, we get
\begin{equation}
W_{\mag}\sim{\cal M}^{ij}q_iq_j+\frac{\mu}{\wt{g}^2} m_{ij}{\cal M}^{ij}.
\end{equation}
The theory flows down to ${\cal N}=1$ supersymmetric $SO(2N_f-N_c+4)$
QCD with $2N_f$ massless flavors interacting with a meson field
${\cal M}^{ij}$, which is the magnetic theory
in the N.Seiberg's duality \cite{S,IS}.

\section{Duality in $USp(2N_c)$ Gauge Theory}\label{USp}

\subsection{S-duality in ${\cal N}=2$ theory}
In this subsection, we consider ${\cal N}=2$ supersymmetric 
$USp(2N_c)$ QCD
with $2N_c+2$ hypermultiplets in the ${\bf 2N_c}$ representation
of the gauge group.
The superpotential is
\begin{equation}
  W_{\ele}=\sq gQ_a^i\Phi^a_b\J^{bc}Q_c^i+\sq m_{ij}Q_a^i\J^{ab}Q_b^j,
\end{equation}
where 
$a,b=1,\cdots,2N_c$ are color indices, $i,j=1,\cdots,4N_c+4$ are
flavor indices, and
$m=(m_{ij})\equiv({0 \atop 1}{-1 \atop 0})
\otimes\diag(m_1,\cdots,m_{2N_c+2})$
is an anti-symmetric matrix. 

The hyperelliptic curve for the theory was determined in \cite{APSh1}
and was found to be invariant under $T:\tau\ra\tau+1$,
$\prod m_j\ra -\prod m_j$, and $ST^2S:\tau\ra\tau/(1-2\tau)$.
Although it is not a simple strong-weak duality,
we expect that there is a  magnetic theory,
which is also an ${\cal N}=2$ supersymmetric $USp(2N_c)$ QCD
with $2N_c+2$ hypermultiplets in the ${\bf 2N_c}$ representation
of the gauge group,
whose superpotential is
\begin{equation}
  W_{\mag}=\sq\wt{g}q^a_i\varphi_a^b\J_{bc}q^c_i+\sq m^{ij}q^a_i\J_{ab}q^b_j,
\end{equation}
where $\wt{g}$ is the dual gauge coupling in $\wt{\tau}=\tau/(1-2\tau)$.

\subsection{${\cal N}$=1 deformed electric theory}
As we have investigated in $SU$ and $SO$ case,
we break ${\cal N}=2$ supersymmetry to ${\cal N}=1$ supersymmetry
by adding the adjoint mass term:
\begin{equation}
  W_{\ele}=\sq gQ_a^i\Phi^a_b\J^{bc}Q_c^i
  +\sq m_{ij}Q_a^i\J^{ab}Q_b^j+
  \frac{\mu}{\sq}\tr\Phi^2.
  \label{WeleUSp}
\end{equation}
The F-term equations are
\begin{eqnarray}
  gQ_a^iQ_c^i+\mu\J_{ab}\Phi_c^b&=&0,
  \label{phiUSp}\\
  gQ_a^i\Phi^a_b-m_{ij}Q^j_b&=&0.
\end{eqnarray}
In the $USp$ theory, there is no baryon and
the vacuum moduli space is parameterized by 
the meson $M^{ij}\equiv Q_a^i\J^{ab}Q_b^j$.

The constraints for the meson are
\begin{eqnarray}
  M\cdot\wt{M}&=&0, \label{MM''0Sp}\\
  \epsilon_{i_1\cdots i_{4N_c+4}}M^{i_1i_2}\cdots
  M^{i_{2N_c+1}i_{2N_c+2}}&=&0, \label{rankk}
\end{eqnarray}
where we have defined $g^2\wt{M}\equiv g^2M+\mu m$.

(\ref{MM''0Sp}) implies 
\begin{eqnarray}
  M\cdot m &=& m\cdot M. \label{mmmm}
\end{eqnarray}
We set $m_{ij}=\left({0\atop 1}{-1\atop 0}\right)
\otimes\diag(0,\cdots,0,m_{N_f+1},\cdots,m_{N_c-2})$, 
where $m_i$'s are chosen to be generic, and then
(\ref{mmmm}) implies
\begin{eqnarray}
  g^2M &=& 
  \left(
    \begin{array}{cc|cccccc}
      X&0&Y&0\\
      0&0&0&d \\
      \hline
      -Y^T&0&\hs{1.5ex}Z\hs{1.5ex}&0 \\
      0&-d&0&0
    \end{array}
  \right),
\end{eqnarray}
where $X$ and $Z$ are $N_f \times N_f$ anti-symmetric matrices, $Y$ is an
$N_f \times N_f$ matrix and $d=\diag (d_{N_f+1},\cdots
,d_{2N_c+2})$. From (\ref{MM''0Sp}) we find,
\begin{eqnarray}
  \left(
    \begin{array}{cccccccc}
      X&Y\\
      -Y^T&Z
    \end{array}
  \right)\cdot
    \left(
    \begin{array}{cccccccc}
      X&Y\\
      -Y^T&Z
    \end{array}
  \right)=0, \label{abcd}\\
  d_i(d_i-\mu m_i)=0,~~~i=N_f+1,\ldots ,2N_c+2
\end{eqnarray}
 Using similarity transformation of the massless flavor
symmetry group $O(2N_f)$, the solutions can be reduced to $X=Z=0$ and 
\begin{eqnarray}
  Y&=&
  \left(
    \begin{array}{cccccc}
      -\hat{q}^2&\hs{-1ex}-i\hat{q}^2 \\
      \hs{-1ex}-i\hat{q}^2&\hs{1ex}\hat{q}^2\\
      &&0
    \end{array}
  \right),
  \label{USpB} \\
  &&\hat{q}=\diag (q_1,\cdots ,q_r),~~~q_i \in \R_+, \nn\\
  d&=&\diag (\mu m_{N_f+1},\cdots ,\mu m_{N_f+s},0,\cdots ,0,),
  \label{USpd}\\
  &&~~~~~~s\leq 2N_c+2-N_f.\nn
\end{eqnarray}
(\ref{rankk}) implies $r+s\leq N_c$. We can check that this form is
exactly the same as that 
derived from the D-term and F-term equations for $Q$ and $\Phi$ (see
appendix \ref{sp}), and so the constraints (\ref{MM''0Sp}) and
(\ref{rankk}) form a complete set.

Next we consider quantum effects. When the meson is as above,
 the theory turns out to be
${\cal N}=1$ supersymmetric $USp(2(N_c-r-s))$ QCD with $(N_f-2r)$
massless flavors in the defining representation.
For $N_f-2r\leq N_c-r-s$ Affleck-Dine-Seiberg type superpotential is
generated non-perturbatively, and the classical vacua are lifted \cite{IP}.
For $N_f-2r=N_c-r-s+1$ the theory is in the confining phase and
the classical moduli space is
deformed quantum mechanically \cite{IP}.
We will consider the branches which have a point $Y=0$ in 
(\ref{USpB}), and discuss the duality on these branches.
Thus we find a constraint $N_c-N_f+r-s+1<0$.

\subsection{${\cal N}$=1 deformed magnetic theory and duality}
The superpotential of ${\cal N}=1$ deformed magnetic theory is
\begin{eqnarray}
  W_{\mag}=\sq \wt{g}q^a_i\varphi_a^b\J_{bc} q^c_i 
  +\sq m^{ij}q^a_i\J_{ab}q^b_j-\frac{\mu}{\sq}\tr\varphi^2.
\end{eqnarray}
The meson in the magnetic theory is defined as 
$N^{ij}=N_{ij}\equiv q^a_i\J_{ab}q^b_j$
which is constrained by the similar equation $N\cdot\wt{N}=0$,
where $\wt{g}^2\wt{N}\equiv \wt{g}^2N-\mu m$, and $\rank N\leq 2N_c$.

The correspondence between the electric and magnetic theory is
\begin{eqnarray}
  \mbox{electric}&\lra& \mbox{magnetic} \nn \\
  g^2M&\lra&\wt{g}^2\wt{N} \\
  (g^2\wt{M}&\lra&\wt{g}^2N). \nn 
\end{eqnarray}
The correspondence of the constraint (\ref{MM''0Sp}) is trivial:
\begin{equation}
  M\cdot\wt{M}=0\lra N\cdot \wt{N}=0.
\end{equation}
When $M$ takes the form as in (\ref{USpB}) and (\ref{USpd}),
we find $\rank N=2(2N_c-N_f+r-s+2)$. As explained above, the vacuum
moduli space is constrained with $N_c-N_f+r-s+1<0$ and so we find
$\rank N\leq 2N_c$. These facts imply that the vacuum moduli spaces
in the electric and magnetic theory are the same. 

We can determine the unbroken gauge groups in the same way as in
section \ref{SOgauge}.
(\ref{phiUSp}) and the similar equation in the magnetic theory imply that
\begin{equation}
  \tr\Phi^k=\left(-\frac{g}{\mu}\right)^k\Tr M^k,~~~
  \tr\varphi^k=\left(\frac{\wt{g}}{\mu}\right)^k\Tr N^k.
  \label{phibyMUSp} 
\end{equation}
The point $M=0$ in the electric theory at which
$USp(2N_c)$ gauge symmetry is expected to be unbroken,
corresponds to the point $\wt{g}^2N=\mu m$.
At this point, (\ref{phibyMUSp}) implies that
\begin{equation}
  \wt{g}\varphi=
  \left(\begin{array}{cc}
      i&0\\
      0&-i \end{array}\right)
  \otimes
  \left(\begin{array}{cccccc}
      0&       &  &       &       &        \\
      &\ddots &  &       &       &        \\
      &       &0 &       &       &        \\
      &       &  &\mms{m_{N_f+1}}&     &        \\
      &       &  &       &\ddots &        \\
      &       &  &       &       &m_{2N_c+2}
    \end{array}\right).
\end{equation}
So we expect that the magnetic theory at this point is a
$USp(2(N_f-N_c-2))$ gauge theory.
The result is consistent to N.Seiberg's
duality in ${\cal N}=1$ $USp$ QCD \cite{S,IP}.
Similarly, when we consider the branch of $\rank d=s$ in (\ref{USpd})
in the electric theory, we get the duality for $USp(2(N_c-s))\lra
USp(2(N_f-N_c+s-2))$. Thus we have found  a series of $USp$ duality as in the
case of the $SO$ theory.

We can also carry out Leigh-Strassler transformation
in this case. Integrating out $\Phi$ in (\ref{WeleUSp}),
\begin{equation}
  W_{\ele}=-\frac{g^2}{\sq\mu}(Q^i\J Q^j)(Q^j\J Q^i)
  +\sq m_{ij}(Q^i\J Q^j).
\end{equation}
In the limit $g\ra 0$, we have
\begin{equation}
  W_{\ele}\sim \sq m_{ij}(Q^i\J Q^j).
\end{equation}
The theory is ${\cal N}=1$ supersymmetric $USp(2N_c)$ QCD with
$N_f$ massless flavors.

On the other hand, introducing suitable auxiliary meson field
${\cal M}^{ij}$, the superpotential for the magnetic theory becomes
\begin{equation}
  W_{\mag}=-{\cal M}^{ij}(q_i\J q_j)+\frac{\mu}{\wt{g}^2}m^{ij}
  {\cal M}^{ij}+\frac{\mu}{2\sq\wt{g}^2}{\cal M}^{ij}{\cal M}^{ij}.
\end{equation}
The equation of motion implies
$\mu{\cal M}^{ij}=\sq(\wt{g}^2N^{ij}-\mu m^{ij})\lra\sq g^2 M^{ij}$.
In the same limit 
$\wt{g}\sim 1/g \ra \infty$ , fixing $\VEV{q\J q}\sim \mu m/\wt{g}^2$
and $\VEV{\varphi}\sim m/\wt{g}$ , we get
\begin{equation}
  W_{\mag}\sim-{\cal M}^{ij}(q_i\J q_j)+\frac{\mu}{\wt{g}^2}m^{ij}
  {\cal M}^{ij}.
\end{equation}
The theory again flows down to the magnetic theory of N.Seiberg's
duality, namely ${\cal N}=1$ supersymmetric $USp(2(N_f-N_c-2))$ QCD with
$N_f$ massless flavors interacting with a meson field ${\cal M}^{ij}$
\cite{S,IP}.

\section{Summary and Comments}
We have studied the deformations of ${\cal N}=2$ supersymmetric QCD,
adding the adjoint mass term, and seen that
the S-dualities in finite ${\cal N}=2$ theories naturally flow to N.Seiberg's
${\cal N}=1$ dualities. Generalizing the argument in \cite{LS}, we
obtained the gauge singlet meson field and the superpotential needed
in the magnetic theory.

We have shown that much information on the duality can be obtained
using the classical equations of motion. This fact would be the remnant
of the non-renormalization
theorem of the Higgs branches in ${\cal N}=2$ theories.
Of course, it is important to take
the non-perturbative effects into account, when we break ${\cal N}=2$
supersymmetry to ${\cal N}=1$. In particular, in order to show that
the vacuum moduli spaces in the electric and magnetic theory are the
same, the non-perturbative effects are essential.
However, we have not revealed more fruitful structures in the ${\cal N}=1$
theories, such as the confining phases, quantum deformations of the
moduli spaces, or electric-magnetic-dyonic triality in the $SO$ theory
\cite{lec,IS}. 

Our methods are quite simple and seem to have other applications.
For example, instead of adding the adjoint mass term, we can add
various ${\cal N}=2$ breaking terms to the superpotential and
investigate deformations of the S-dualities.
We hope that this approach will give a useful guide for searching new
dualities. 

\section*{Acknowledgments}
We would like to thank our colleagues in Kyoto University for valuable 
discussions and encouragement. We are especially grateful to T.Harano
for various discussions about duality.
The work of T.H. and S.S. is supported in part by the Grant-in-Aid
for JSPS fellows. The work of N.M is supported in part by the
Grant-in-Aid for Scientific Research from the Ministry of Education,
Science and Culture.


\section*{Appendices}
\appendix

\section{convention} \label{conv}
We write down the conventions used in the $SO(N_c)$ gauge theory.
\begin{eqnarray*}
  &\bullet&
  \J=(\J_{ij})\equiv \left({0\atop -1}{1\atop 0}\right)\otimes\I
  \\
  &\bullet&
  \J^{-1}=(\J^{ij})\equiv \left({0\atop 1}{-1\atop 0}\right)\otimes\I
  \\
  &\bullet&
  \J_{ij}\J^{jk}=\J^{kj}\J_{ji}=\delta_i^k
  \\
  &\bullet&
  m_{ij}=\left({0\atop 1}{1\atop 0}\right)\otimes\diag(m_1,\cdots,m_{N})
  \\
  &\bullet&
  m^{ij}\equiv\J^{ik}m_{kl}\J^{lj}=(\J m\J)^{ij}=m_{ij}
  \\
  &\bullet&
  m^{ij}\J_{jk}=\J^{ij}m_{jk}
  \\
  &\bullet&
  N_{ij}=q^a_iq^a_j
  \\
  &\bullet&
  N^{ij}\equiv\J^{ik}N_{kl}\J^{lj}=(\J N\J)^{ij}
\end{eqnarray*}

\section{the form of Q} \label{Qform}
We list the D-term and F-term
equations and the classical moduli spaces in terms of $Q,\wt{Q}$ and $\Phi$.

\subsection{${\cal N}=2~SU$ theory} \label{SU}
The D-term and F-term equations are
\begin{eqnarray}
  [\Phi, \Phi^{\dag}] &=& 0, \label{dd1}\\
  Q^{\dag a}_i Q^i_b - \wt{Q}^a_i \wt{Q}^{\dag i}_b &=& \nu \delta_b^a 
  , \label{dd2}\\
  Q^i_a\wt{Q}^b_i &=& \rho \delta_a^b,\label{ff1} \\
  g\Phi^a_b\wt{Q}^b_j +m^i_j\wt{Q}^a_i &=& 0, \label{ff2}\\
  gQ^i_a\Phi^a_b +m^i_jQ^j_b &=& 0 \label{ff3}, 
\end{eqnarray}
where $\nu \in \R$, $\rho \in \C$ and $m^i_j=\diag(0,\cdots ,0, m_{N_f+1},
\cdots, m_{2N_c})$
\footnote{
Using the F-term equations (\ref{ff2}) and (\ref{ff3}), we find
\begin{eqnarray*}
D^2&=&g^2\left( |Q^{\dag} T^{\alpha}
  Q-\wt{Q}T^{\alpha}\wt{Q}^{\dag}|^2 + 2 \tr
  ([\Phi^{\dag},\Phi]^2)
  +4( |Q^{\dag} m+Q^{\dag}\Phi|^2
    +|m^{\dag} \wt{Q}+\Phi^{\dag}\wt{Q}|^2) \right),
\end{eqnarray*}
implying (\ref{dd1}) and (\ref{dd2}).
}.
The vacuum moduli space consists of the Coulomb, baryonic
and non-baryonic branches. 

\paragraph{The Coulomb branch}
$Q$ and $\wt{Q}$ are zero.
\begin{eqnarray}
  \Phi &=&
  \left(
    \begin{array}{ccccccccccc}
      \phi_1 \\
      &\ddots \\
      &&\phi_{N_c} \\
    \end{array}
  \right),
\end{eqnarray}
where $\phi_i \in \C$ and $\sum \phi_i=0$.

\paragraph{The Baryonic Branch}
$\Phi$ is zero.
\begin{eqnarray}
  Q &=&
  \left(
    \begin{array}{ccccccccccc}
      \kappa_1 \\
      &\ddots \\
      &&\ddots \\
      &&&\kappa_{N_c}\hs{24ex}
    \end{array}
  \right), \\
  \wt{Q} &=& 
  \left(
    \begin{array}{ccccccccccc}
      \wt{\kappa}_1&&&&\lambda_1 \\
      &\ddots &&&&\ddots \\
      &&\ddots &&&&\lambda_{N_f-N_c}\hs{6ex} \\
      &&&\wt{\kappa}_{N_c}
    \end{array}
  \right),
\end{eqnarray}
\begin{eqnarray*}
 {\rm where}& \kappa_i, \lambda_i \in \R_+, \\
 &\kappa_i\wt{\kappa}_i =\rho, & \mbox{ for all } i,\\
 &\kappa^2_i-(|\wt{\kappa}_i|^2 + \lambda^2_i) = \nu, & i\leq N_f-N_c, \\
 &\kappa_i^2 -|\wt{\kappa}_i|^2 = \nu, & i\geq N_f-N_c+1.
\end{eqnarray*}

\paragraph{The non-Baryonic Branch}
\begin{eqnarray}
\Phi&=&\diag (0,\cdots,0,\phi_{r+1},\cdots,\phi_{N_c}),\\
  Q &=& 
  \left(
    \begin{array}{ccccccccccc}
      \kappa_1 \\
      &\ddots \\
      &&\kappa_r\hs{24ex} \\
      \\
    \end{array}
  \right),\\
  \wt{Q}&=& 
  \left(
    \begin{array}{ccccccccccc}
      0 &&&\kappa_1\\
      &\ddots &&&\ddots  \\
      &&0 &&&\kappa_r\hs{14ex} \\
      \\
    \end{array}
  \right),
\end{eqnarray}
where $\phi_i\in \C$ , $\sum \phi_i =0$ , $\kappa_i \in \R_+$ and $r
\leq [N_f/2]$ .

\subsection{${\cal N}=1$ deformed $SU$ theory}\label{SU2}
The D-term equations (\ref{dd1}) and (\ref{dd2}) are not
changed when we add the adjoint mass term $\mu \tr \Phi^2/\sqrt{2}$.

The D-term and F-term equations are
\begin{eqnarray}
  [\Phi, \Phi^{\dag}] &=& 0, \label{d1}\\
  Q^{\dag a}_i Q^i_b - \wt{Q}^a_i \wt{Q}^{\dag i}_b &=& \nu ,\label{d2}\\
  g\Big( Q^i_a\wt{Q}^b_i -\frac{1}{N_c}(Q^i\wt{Q}_i)\delta^b_a \Big) +
  \mu \Phi^b_a &=& 0, \label{d3}\\
  g\Phi^a_b\wt{Q}^b_j +m^i_j\wt{Q}^a_i &=& 0, \label{d4}\\
  gQ^i_a\Phi^a_b +m^i_jQ^j_b &=& 0, \label{d5}
\end{eqnarray}
where $\nu \in \R_+$ and $m^i_j=\diag(0,\cdots ,0, m_{N_f+1}, \cdots,
m_{2N_c})$.

The classical moduli space consists of the baryonic , non-baryonic 
and exceptional branches. The baryonic and non-baryonic
branches are the same as those in ${\cal N}=2$ theory ($\Phi$ is fixed to
be zero).

\paragraph{The Exceptional Branch}
\begin{eqnarray}
  g\Phi&=&\diag(0, \cdots, 0, \underbrace{c, \cdots , c}_{N_c-r-s},
  -m_{2N_c-s+1}, \cdots, -m_{2N_c}), \\
  Q &=&
  \left(
    \begin{array}{ccccccccccc}
      \kappa_1 \\
      &\ddots \\
      &&\kappa_r \hs{19ex} \\
      \\     
      &&&\mms{d_{N_f+1}} \\
      &&&&\ddots \\
      &&&&&\mms{d_{N_f+s}}\hs{6ex}\\\
      \\
    \end{array}
  \right), \\
  \wt{Q}&=& 
  \left(
    \begin{array}{ccccccccccc}
      \wt{\kappa_1} &&&\lambda_1\\
      &\ddots &&&\ddots\\
      &&\raisebox{-1ex}{$\wt{\kappa_r}$}&
      &&\raisebox{.5ex}{$\lambda_{r^{'}}$}\hs{6ex}
\\
      \\      
      &&&&&&\mms{\wt{d}_{N_f+1}} \\
      &&&&&&&\ddots \\
      &&&&&&&&\mms{\wt{d}_{N_f+s}}\hs{6ex}\\
      \\
    \end{array}
  \right),
\end{eqnarray}
\begin{eqnarray*}
  {\rm where}&& (\kappa_i, d_i, \lambda_i) \in \R_+ ,~~ (\wt{\kappa}_i,
  \wt{d}_i) \in \C, \\
  && r+s\leq N_c-1,~~ s\leq 2N_c-N_f,~~ r'=\min\{r, N_f-r\}, \\
  && c = \frac{1}{N_c-r-s}\sum_{2N_c-s+1}^{2N_c}m_i, \\
  && g^2\kappa_i\wt{\kappa}_i=\mu c,~~ \kappa_i^2
  -(|\wt{\kappa}_i|^2+\lambda_i^2)=0, \\
  && d_i=|\wt{d}_i|,~~ g^2 d_i\wt{d}_i = \mu c +\mu m_{i}~~(N_f+1\leq
  i \leq N_f+s).
\end{eqnarray*}

\subsection{${\cal N}=1$ deformed $SO$ theory} \label{so}
As in the $SU$ theory,
the D-term equations are not changed from ${\cal N}=2$ theory.
The D-term and F-term equations are
\begin{eqnarray}
  [\Phi,\Phi^{\dag}]&=& 0, \label{so1}\\
  \Im (Q^{\dag}_{ia}Q^i_b) &=& 0, \label{so2}\\
  gQ^i_a\J_{ij}Q^j_b-\mu\Phi_{ab} &=&0, \label{so3}\\
  g\Phi_{ab}Q^j_b\J_{ij} +m_{ij}Q^j_a &=&0, \label{so4}
\end{eqnarray}
where $m=(m_{ij})\equiv({0\atop 1}{1\atop
  0})\otimes\diag(0,\cdots, 0, m_{N_f+1}, \cdots , m_{N_c-2})$. 

The classical moduli space is as follows.
\begin{eqnarray}
  g\Phi &=&
  \left(
    \begin{array}{ccccccccccccccc}
      0\\
      &\ddots \\
      && 0 \\
      &&&&\mms{im_{N_f+1}}\\
      &&&\mms{-im_{N_f+1}}& \\
      &&&&&\ddots \\
      &&&&&&&\mms{im_{N_f+s}}\\
      &&&&&&\mms{-im_{N_f+s}}&\\
      &&&&&&&&0 \\
      &&&&&&&&&\ddots\hs{2ex}
    \end{array}
  \right), \\
  Q &=& 
  \left(
    \begin{array}{ccccccc|cccccccccc}
      q_1&&&&&&&& \\
      &\ddots&&&&&&&& \\
      &&q_r \hs{1ex}&&&&&&& \\
      &&&&&&&&&&\\
      &&&d_1&&&&\hs{9ex}&d_1 \\
      &&&\hs{-2.5ex}\makebox[2.5ex]{$-id_1$}&&&&&\hs{-.7ex}id_1\\
      &&&&\ddots&&&&&\ddots \\
      &&&&&d_s&&&&&d_s \\
      &&&&&\hs{-2.5ex}-id_s&&&&&\hs{-.7ex}id_s \\
      &&&&&&\hs{.7ex}&&&&&\\
    \end{array}
  \right),
\end{eqnarray}
where $q_i \in \R_+$ ,~~$r+2s \leq N_c$ ,~~$r\leq N_f$,~~$s\leq
  N_c-2-N_f$ and $d_i=\sqrt{\frac{\mu m_{N_f+i}}{2g^2}}$.

\subsection{${\cal N}=1$ deformed $USp$ theory}\label{sp}
The D-term and F-term equations are
\begin{eqnarray}
  [\Phi, \Phi^{\dag}] &=& 0, \label{sp1}\\
  Q^{\dag i}_aQ^i_b +Q^{\dag i}_bQ^i_a &=& 0, \label{sp2}\\
  gQ^i_aQ^i_c + \mu \J_{ab}\Phi^b_c &=& 0, \label{sp3}\\
  gQ^i_a\Phi^a_b-m_{ij}Q^j_b &=& 0 \label{sp4}, 
\end{eqnarray}
where $Q^{\dag i}_a \equiv Q^{\dag ib}\J_{ba}$ and
$m=(m_{ij})\equiv({0 \atop 1}{-1 \atop 0})
\otimes\diag(0,\cdots, 0, m_{N_f+1}, \cdots , m_{2N_c+2})$.

The classical moduli space is as follows.
\begin{eqnarray}
  g\Phi&\hs{-1ex}=&\hs{-1.7ex}
  \left(
    \begin{array}{cc}
      1\\
      &-1
    \end{array}
  \right)\otimes \diag(0,\cdots ,0 ,im_{N_f+1},\cdots
  ,im_{N_f+s},0,\cdots ,0), \\
  Q &\hs{-1ex}=&\hs{-1.7ex}
  \left(
    \begin{array}{ccccccccccccccc}
      Q'&0&0&0 \\
      0 &iD&0&-D \\
      0 &0&Q'&0 \\
      0&-D&0&iD
    \end{array}
  \right), \\
  \raisebox{6ex}{where}&&Q'=
  \left(
    \begin{array}{ccccccccccccccc}
      q_1&&&iq_1 \\
      &\ddots&&&\ddots \\
      &&q_r&&&iq_r \hs{4ex}\\
      \\
    \end{array}
  \right),\\
  &&~~~~~Q' {\rm~is~an~} (N_c-s)\times (N_f) {\rm~matrix},\nn \\
  &&~~~~~r\leq {\rm min}\{N_c-s,[N_f/2]\},~~q_i \in \R_+,\nn\\
  &&gD=
  \left(
    \begin{array}{cccc}
      \sqrt{\mu m_{N_f+1}/2}\\
      &\ddots \\
      &&\sqrt{\mu m_{N_f+s}/2}\hs{4ex}
    \end{array}
  \right),\\
  &&~~~~~s\leq 2N_c+2-N_f,~~r+s \leq N_c.\nn
\end{eqnarray}


\end{document}